\documentclass[aps,prapplied,superscriptaddress]{revtex4-2}
\usepackage[utf8]{inputenc}
\usepackage[T1]{fontenc}
\usepackage{textcomp, mathtools, amssymb, amsthm, wasysym, gensymb}
\usepackage[colorlinks=true, linkcolor=red, citecolor=blue, urlcolor=blue]{hyperref}

\usepackage{pdfpages,pgffor}
\makeatletter
\AtBeginDocument{\let\LS@rot\@undefined}
\makeatother

\newcommand\unine{Laboratoire Temps-Fréquence, Institut de Physique, Université de Neuchâtel, Avenue de Bellevaux 51, 2000 Neuchâtel, Switzerland}
\newcommand\ethz{Institute for Quantum Electronics, ETH Zurich, Auguste-Piccard-Hof 1, 8093 Zurich, Switzerland}

\begin{document}

\title{Free-Running Ring Quantum Cascade Laser with 50~kHz Linewidth}

\affiliation{\unine}
\affiliation{\ethz}

\author{Alexandre~Parriaux}
\email[Corresponding author: ]{alexandre.parriaux@unine.ch}
\affiliation{\unine}

\author{Ina~Heckelmann}
\affiliation{\ethz}

\author{Mathieu~Bertrand}
\affiliation{\ethz}

\author{Mattias~Beck}
\affiliation{\ethz}

\author{Jérôme~Faist}
\affiliation{\ethz}

\author{Thomas~Südmeyer}
\affiliation{\unine}

\date{\today}

\begin{abstract}
We report on the noise characterization of a free-running ring quantum cascade laser resonator emitting a single frequency mode around 7.7~µm. Using a gas cell filled with N$_2$O as a frequency-to-voltage discriminator, we measured the frequency noise power spectral density of the laser from which we extracted its linewidth. The results show a full width at half maximum close to 50~kHz at 1~s integration time, which represents at least a sixfold improvement compared to state-of-the-art quantum cascade lasers operating in a spectral region above 7~µm.
We also demonstrate that such lasers can be efficiently used for frequency modulation spectroscopy, which opens up new possibilities for high resolution metrology and spectroscopic applications in the mid-infrared.
\end{abstract}

\maketitle

\section{Introduction}
Single frequency mode lasers operating in the mid-infrared (MIR) above 7~µm are important tools for many applications such as spectroscopy~\cite{komagata-oe-2022}, atmospheric measurements~\cite{damato-oe-2025}, ranging~\cite{martin-jlt-2025}, biomedicine~\cite{schwaighofer-chemsocrev-2017,klocke-analchem-2018}, but also for fundamental physics~\cite{argence-natphot-2015,santagata-optica-2019}. Lasers operating in this spectral region are either based on nonlinear frequency conversion of near-infrared sources, or based on quantum cascade lasers (QCL)~\cite{faist-science-1994,botez-book-2023}.
Compared to experimental setups based on frequency conversion, QCLs have the advantage of being more compact, able to generate high output power~\cite{razeghi-lsa-2025}, and with strong potential for large-scale production. However, QCLs are also known to be sensitive to noise~\cite{Schilt-apb-2015,gabbrielli-oe-2025,lapenna-pra-2026}, which prevents their use in several applications.
Low-noise operation of QCLs can nevertheless be achieved using various stabilization techniques~\cite{argence-natphot-2015,santagata-optica-2019,shehzad-ol-2019,sinhal-molphys-2022}, but the required experimental setups are often bulky and expensive. Hence, stabilizing QCLs is not a practical solution to enable widespread use of such lasers in embedded systems, typically for out-of-the-laboratory experiments.
To the best of our knowledge, in the spectral region above 7~µm, the narrowest linewidth measured at 1~s integration time using a QCL operating in a free-running mode was close to 300~kHz as reported in Ref.~\cite{chomet-apl-2023}.
In contrast, around 1550 nm, single frequency mode lasers with linewidths below 10~kHz at 1~s integration time are readily commercially available.

In this letter, we characterized the noise properties of a free-running ring QCL resonator operating around 7.7~µm. This device, which has already shown enhanced properties when operating in a frequency comb regime~\cite{heckelmann-science-2023}, also shows improved performance in free-running compared to state-of-the-art single frequency mode lasers emitting in a similar spectral region. Using a frequency-to-voltage discriminator, the laser's linewidth at 1~s integration time was measured to be close to 50~kHz, which represents at least a sixfold improvement compared to the state-of-the-art.

\section{Experimental setup}

\begin{figure}[t]
    \centering
    \includegraphics{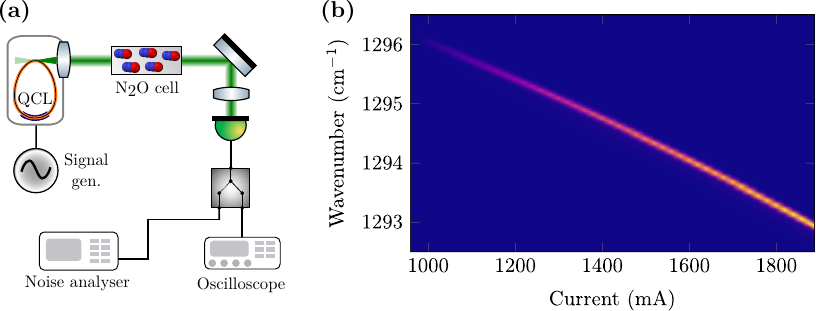}
    \caption{\textbf{(a)} Experimental setup used to characterize the noise properties of a ring QCL using a N$_2$O gas cell as a frequency-to-voltage discriminator. \textbf{(b)} Spectral tuning range of the laser with respect to the current supplied, when operating at a temperature of 0\celsius.}
    \label{fig:setup}
\end{figure}

The experimental setup considered here is schematized in \autoref{fig:setup}~\textbf{(a)}. The ring QCL being characterized has a structure that has already been reported in Ref.~\cite{heckelmann-science-2023} as ``Device~1''. The materials, fabrication and egg-shaped cavity design details can be found in the supplementary materials of Ref.~\cite{heckelmann-science-2023}, and in Ref.~\cite{beck-science-2002} for the buried heterostructure process.
The temperature of the laser is set to 2\celsius{} and the current to 1515~mA, both parameters being controlled by a low-noise laser driver (ppqSense, QubeCL20-T). Under these conditions, the QCL emits a stable single-frequency mode signal close to 1294~cm$^{-1}$.
The spectral tuning range of the central frequency of the QCL, when operating at 0\celsius, is presented in \autoref{fig:setup}~\textbf{(b)} which shows a set of recorded optical spectra with a Fourier transform infrared spectrometer at different driving currents applied.

The optical output beam of the QCL is collimated with a 4~mm focal length lens, and passes either through a 10~cm or 1~cm long gas cell, both filled with 1~mbar of N$_2$O gas. The 10~cm gas cell is used for frequency calibration purposes, whereas the 1~cm gas cell is used as a frequency-to-voltage discriminator (see below).
At the output of the gas cell, the beam is focused on a photodetector, and the detector's output is either directed towards an oscilloscope or a phase noise analyzer (Rohde \& Schwarz, FSWP26).

To characterize the noise properties of the QCL and especially its frequency noise power spectral density (FN-PSD) profile, we use a frequency-to-voltage discriminator to convert laser frequency fluctuations to measurable voltage fluctuations of the photodetector. For this, we use the 1~cm long gas cell, and especially the linear part of one side of the N$_2$O absorption line centered at 1293.89~cm$^{-1}$ (R(10) line of the fundamental $\nu_1$ band).
This absorption line has to be analyzed first to measure the discriminator's value, i.e., the factor to convert voltage to frequency fluctuations, which also means that our experimental setup has to be frequency calibrated.
Such a calibration is performed by applying a 3.5~V peak ramp to the input modulation port of the laser driver using a signal generator, which corresponds to a 35~mA peak modulation of the laser current around its setpoint. To observe more lines, which is beneficial for calibration purposes, we used the 10~cm long cell instead of the 1~cm one. The electrical output of the detector is directed towards the oscilloscope, which is triggered on the signal generator.
With the 10~cm long cell, we can observe several absorption lines on the oscilloscope between 1293.98~cm$^{-1}$ and 1294.11~cm$^{-1}$, and the central frequency of these lines, which we got from the HITRAN database~\cite{gordon-jqrst-2022}, are then used as frequency ticks to convert the time axis of the oscilloscope to an optical frequency axis.

\section{Results}

\begin{figure}[t]
    \centering
    \includegraphics{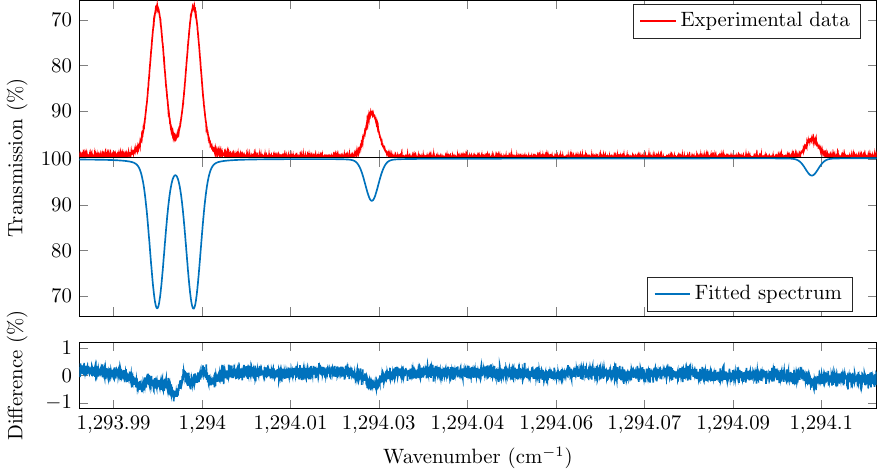}
    \caption{\textbf{(Top)} Transmission spectrum observed with a 10~cm long gas cell filled with 1~mbar of N$_2$O when current modulation is on. \textbf{(Middle)} Fitted absorption spectrum using line parameters from the HITRAN database. \textbf{(Bottom)} Absolute difference between the experimental and fitted spectra.} 
    \label{fig:spectro}
\end{figure}

Frequency modulation spectroscopy on N$_2$O was performed not only to demonstrate the suitability of the ring QCL for spectroscopic applications, but also to establish a frequency calibration for the future measurements.
The experimentally measured transmission spectrum (\autoref{fig:spectro}~(top)) was fitted with a set of Voigt profiles whose line parameters were obtained from the HITRAN database~\cite{gordon-jqrst-2022} (\autoref{fig:spectro}~(middle)). 
The absolute difference between the experimental and fitted spectrum (\autoref{fig:spectro}~(bottom)) shows a high degree of agreement.

\begin{figure}[t]
    \centering
    \includegraphics{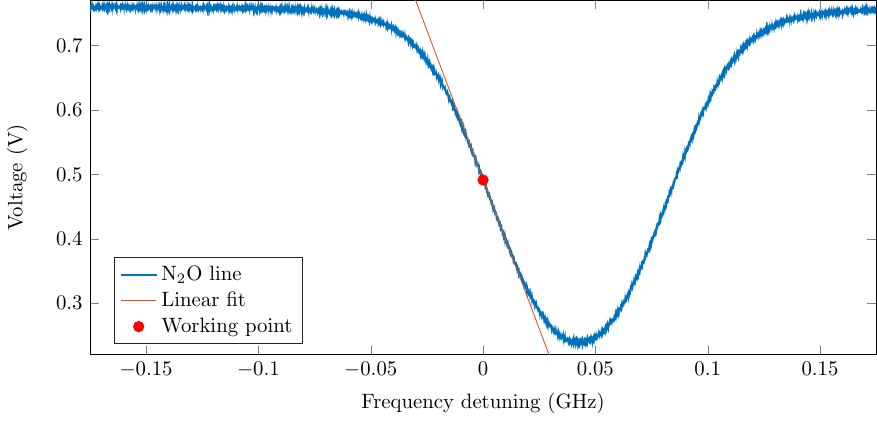}
    \caption{R(10) absorption line of the $\nu_1$ fundamental band of N$_2$O measured by performing current modulation of the QCL. The red line aims to show that at the vicinity of the working point, the absorption line has a linear behavior.}
    \label{fig:freq_discr}
\end{figure}

Once the calibration procedure is done, we swap the 10~cm long gas cell with the 1~cm one and slightly modify the current of the laser to set its frequency at a working point allowing us to operate in the linear part of the R(10) absorption line, as can be seen in \autoref{fig:freq_discr}. We then performed current modulation again, with a smaller peak amplitude compared to the calibration procedure, which allows us to observe the absorption line. After converting the time axis of the oscilloscope to frequency axis, we numerically extracted the frequency-to-voltage discriminator's value, i.e., the value of the slope of the tangent at the working point, which is equal to $D = -9.26 $~V/GHz.
\autoref{fig:freq_discr} also shows the tangent line at the working point to check that, at the vicinity of the working point, the absorption line has a linear behavior.

\begin{figure}[t]
    \centering
    \includegraphics{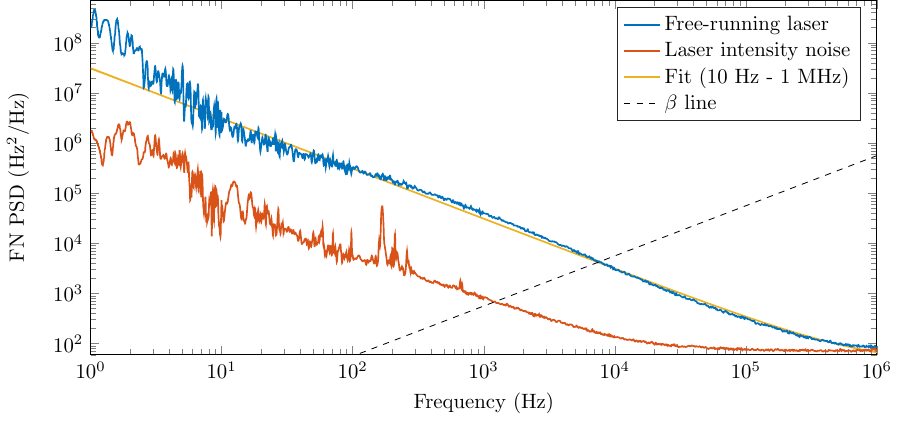}
    \caption{FN-PSD measurement obtained by converting back the voltage noise to frequency noise using the discriminator's value. The red curve shows that the intensity noise of the QCL does not contribute to the frequency noise.} 
    \label{fig:fnpsd}
\end{figure}

When current modulation is off, and the laser operates at the working point showed in \autoref{fig:freq_discr}, the output of the detector is directed to the phase noise analyzer to measure its voltage noise spectral density. Using the discriminator's value, it is converted to FN-PSD and the results are presented in \autoref{fig:fnpsd} (blue curve). The same measurement is performed after slightly changing the current of the laser (and hence its central frequency) to be outside any absorption feature of the gas.
This is done to evaluate the contribution of the intensity noise of the laser to the FN-PSD measurement. As can be seen in \autoref{fig:fnpsd} (red curve), the intensity noise contribution is well below the laser FN-PSD at all frequencies, which allows us to conclude that our FN-PSD measurements are not limited by the intensity noise of the laser.

The FN-PSD of the QCL is mostly dominated by flicker frequency noise at all frequencies, except at high frequencies close to 1~MHz where white frequency noise can be observed but originates from detection limits in our setup. Below 10~Hz, we also see an increase of the frequency noise which we attribute to environmental mechanical/vibrational perturbations that can typically be associated to the chiller used, the dry air flow in the housing of the laser to prevent water condensation, etc.
A comparison of the experimental data in the 10~Hz -- 1~MHz region with a fitted function composed of flicker and white frequency noise defined by:
\begin{equation} \label{eq:fnpsd_typ}
    S_{\nu}(f) = h_0 + \alpha f^{-1} 
\end{equation}
where $\alpha$ is the flicker noise coefficient, and $h_0$ is the white noise coefficient, shows that flicker frequency noise is the main contribution behind the noise properties of the laser.
From the fit, we can extract the noise type coefficients which are found to be $\alpha = 3.1 \times 10^7$~Hz$^2$ and $h_0 = 38$~Hz$^2$/Hz. The latter is a particularly interesting value as it is linked to the intrinsic linewidth $\Delta \nu_\text{Int}$ of the laser via the relation $\Delta \nu_\text{Int} = \pi h_0$.
The obtained value for $h_0$ using the fitting procedure is underestimated as the measured FN-PSD profile only begins to exhibit white noise features at frequencies close 1~MHz.
However, from the experimental data, we can give an upper limit of $h_0$ which we evaluated to be slightly below 100~Hz. This corresponds to an intrinsic linewidth of the QCL of less than 300~Hz. This upper limit is compatible with analytical predictions we derived from Refs.~\cite{Cappelli-optica-2015,meng-natphot-2021} (see Supplementary Materials), which suggest an intrinsic linewidth of less than 10~Hz due to the device's high intracavity power and low total losses.
Compared to Fabry-Perot QCLs emitting a single frequency mode such as in Ref.~\cite{Cappelli-optica-2015}, our ring QCL device shows a higher Q-factor value estimated to be 8200, which represents a 60\% increase (see Supplementary Materials). Consequently, an improved intrinsic linewidth is expected for such ring QCL devices~\cite{jin-natphot-2021}, even if higher output power would be achieved.
Further investigation would be necessary to carefully evaluate experimentally the intrinsic linewidth, but such measurements can be technically demanding as it often requires more sophisticated experimental setups~\cite{Bartalini-prl-2010,Bartalini-oe-2011,Cappelli-optica-2015}.

\begin{figure}[t]
    \centering
    \includegraphics{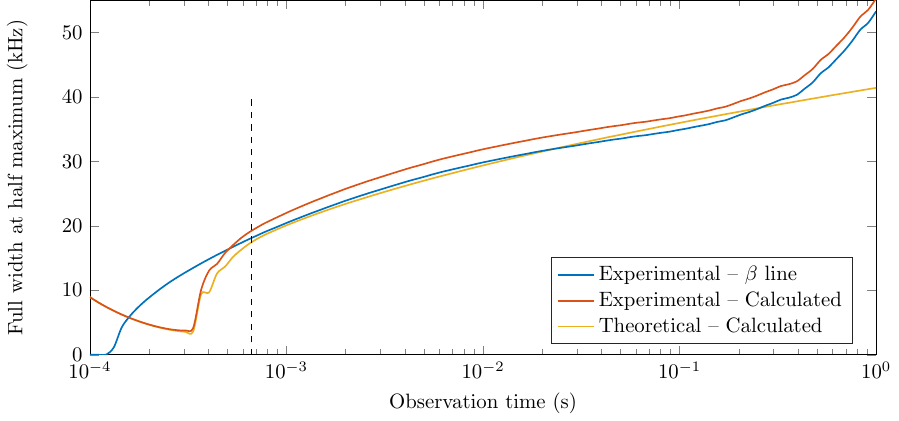}
    \caption{FWHM evolution of the QCL with respect to the observation time, and depending on the method used to extract it. The yellow curve represents the FWHM evolution for a laser having the theoretical FN-PSD profile given by the fitted curve showed in \autoref{fig:fnpsd}. The vertical dashed line indicates the limit below which the $\beta$ line method is not relevant anymore to estimate the linewidth.}
    \label{fig:linewidth}
\end{figure}

Due to the presence of flicker noise, the linewidth of the laser is broadened and depends on the observation time. There are two possibilities to extract the observed linewidth of the QCL using the previous FN-PSD measurements. The first one is to use the well known $\beta$ separation line method~\cite{didomenico-appopt-2010}, which allows to geometrically separate parts of the FN-PSD that contribute either to the linewidth or to the wings of the lineshape.
Using this approach, we extracted the full width at half maximum (FWHM) at different observation times, i.e., the inverse of the minimum frequency at which the FN-PSD is considered. The results are shown in \autoref{fig:linewidth} (blue curve).
The second method is to calculate the lineshape and extract the FWHM. Indeed, it is known that the lineshape of a laser is fully determined by its FN-PSD, and the formula that links the two quantities can be found in Ref.~\cite{elliott-pra-1982}. However, in this original paper, the observation time was not taken into account, which means that the formula used to calculate the lineshape has to be slightly modified as explained in Ref.~\cite{bishof-prl-2013} and in the associated supplementary materials. The FWHM obtained using this method is also shown in \autoref{fig:linewidth} (red curve).

As it is known for the $\beta$ separation line method, in the presence of flicker frequency noise, the linewidth is underestimated by up to 10\%~\cite{didomenico-appopt-2010}, which explains the differences observed when comparing the results obtained with the two methods. Moreover, below an observation time of $T_c = 5/f_\text{cut} \approx 0.7$~ms (showed as a dashed line in \autoref{fig:linewidth}) where $f_\text{cut}$ is the frequency at which the FN-PSD crosses the $\beta$ line, the $\beta$ separation line method is not able to correctly estimate the linewidth.
Hence, below $T_c$, the FWHM can only be extracted from the numerical calculation of the lineshape, and we found that the minimum linewidth of the QCL is achieved at 0.3~ms where is reaches 4~kHz.
When the integration time increases, the FWHM shows a typical increase due to the flicker noise nature of the FN-PSD, and at 1~ms, the FWHM reaches 22~kHz. However, above 0.1~s integration time, the FWHM increases drastically due to the noise excess reported below 10~Hz in \autoref{fig:fnpsd}, and at 1~s integration time, the linewidth reaches a value slightly above 50~kHz.
If the laser would exhibit the same FN-PSD profile as the fitted one showed in \autoref{fig:fnpsd}, the FWHM at 1~s would be close to 40~kHz as can be seen in \autoref{fig:linewidth} (yellow curve) which shows the calculated FWHM evolution for this theoretical FN-PSD profile.

We now compare our findings with state-of-the-art results reported in the literature which is summarized in \autoref{tab:comparison}.
In the spectral region above 7~µm, to the best of our knowledge, the reported results in Ref.~\cite{chomet-apl-2023} showed the free-running QCL with the best noise properties so far. The values reported are 72~kHz and 300~kHz at respectively 1~ms and 1~s integration time. As mentioned above, our results obtained with the ring QCL showed FWHM values of 22~kHz and 50~kHz at respectively 1~ms and 1~s integration time, which corresponds to an improvement of respectively a factor 3 and 6.
Let us note that the QCL used in Ref.~\cite{chomet-apl-2023} is a Fabry-Perot QCL, whereas the other works presented in \autoref{tab:comparison} used a distributed-feedback QCL. Experimentally, the latter is more commonly used to generate a single frequency mode with a QCL as the Fabry-Perot type will start showing a frequency comb structure when pumped at high current levels.
If we compare our results with the ones in \autoref{tab:comparison} obtained with distributed-feedback QCLs, our results show an improvement of a factor 11 (compared to Ref.~\cite{shehzad-ol-2019}) and 8 (compared to Ref.~\cite{manceau-lpr-2025}) respectively at 1~ms and 1~s integration time.

\begin{table}[htb]
    \centering
    \begin{tabular}{|c|c|c|c|} \hline
    \textbf{Reference and year} & \textbf{Wavelength (µm)} & \textbf{FWHM at 1~ms (kHz)} & \textbf{FWHM at 1~s (kHz)}  \\  \hline
    \cite{argence-natphot-2015} (2015) & 10.3 & 300 & 650 \\
    \cite{shehzad-ol-2019} (2019) & 7.8 & 250 & 470 \\
    \cite{chomet-apl-2023} (2023) & 8.1 & 72 & 300 \\
    \cite{manceau-lpr-2025} (2025) & 17 & 290 & 420 \\
    \textbf{This work} & \textbf{7.7} & \textbf{22} & \textbf{50} \\ \hline 
    \end{tabular}
    \caption{Comparison of reported linewidth measurements in the literature for different free-running QCLs emitting a single frequency mode versus the results obtained in this work.}
    \label{tab:comparison}
\end{table}

We believe that the main origin of the reduced flicker frequency noise level observed with our ring QCL (and hence its free-running linewidth) is due to reduced surface state-induced noise~\cite{yamanishi-japp-2014}. Indeed, compared to other QCLs reported in the literature which are based on a straight waveguide architecture, ring QCLs do not present facet-like open surfaces which consequently leads to a reduced number of impurity states. As demonstrated in Ref~\cite{yamanishi-japp-2014}, the induced noise associated with such states are linked to a flicker frequency noise origin which can be reduced by controlling the number of impurity states.

\section{Discussion and conclusion}
In this letter, we have characterized the noise properties of a ring QCL resonator operating in free-running and emitting a single frequency mode around 7.7~µm. Our results showed that such a laser is particularly low-noise, with an observed linewidth at 1~s integration time close to 50~kHz, and with potential to reach 40~kHz if mechanical noises are handled carefully. These results shows a sixfold enhancement compared to state-of-the-art QCL emitting above 7~µm, which we attribute to a reduced number of impurity states and especially to reduced surface state-induced noise~\cite{yamanishi-japp-2014}.
One property we did not comment yet is the output power of our device. In our case, due to the ring geometry, the output power of the QCL was limited to a few hundreds of µW, which can be enough for simple spectroscopic applications as demonstrated here, but limiting for other applications such as ranging~\cite{martin-jlt-2025}. However, recent works have shown that QCL resonators similar as the one considered here can generate higher output power if a passive waveguide is integrated in the design~\cite{cargioli-natcommun-2025,letsou-optica-2026}.
Future works include the noise characterization of these devices and reveal their fundamental linewidth as higher output power will be available.

Although we considered the free-running operation of the ring QCL here, performing radio-frequency injection on this device allows us to generate a quantum walk frequency comb~\cite{heckelmann-science-2023}, and characterizing the performance of such combs would be very interesting as we can expect that the noise properties would also be enhanced compared to Fabry-Perot QCL combs. For this, one possibility would be to use a fully stabilized frequency comb in the mid-infrared as already demonstrated in Ref.~\cite{chomet-optica-2024}, but such measurements can be complex and challenging.

\section*{Acknowledgments}
We acknowledge funding from the Schweizerischer Nationalfonds zur Förderung der Wissenschaftlichen Forschung (Grant No. 40B2-1\_176584 and 200021\_227521); MIRAQLS: Staatssekretariat für Bildung, Forschung und Innovation SBFI (22.00182) in collaboration with EU (grant Agreement 101070700).

\section*{Author Declaration}
The authors have no conflicts to disclose.

\section*{Data availability}
The data that support the findings of this study are openly available in EUDAT B2SHARE at \href{https://doi.org/10.23728/b2share.zre6h-m1474}{10.23728/b2share.zre6h-m1474}~\cite{dataset}.

\bibliography{biblio}

\pagebreak
\foreach \x in {1,2,3}
{%
\clearpage
\includepdf[pages={\x}]{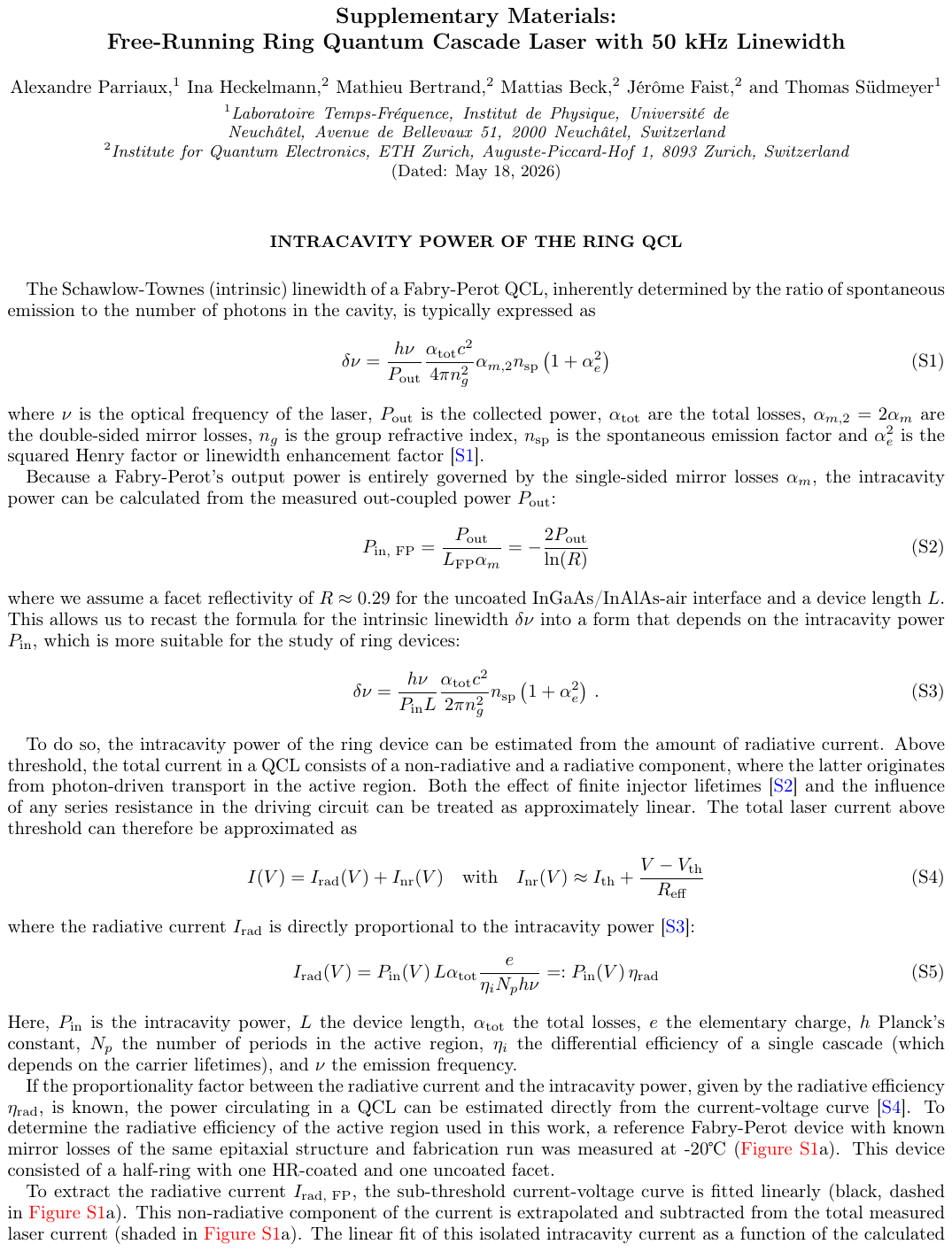}
}

\end{document}